%Paper: cond-mat/9311057
%From: STRINGARI@itnvax.science.unitn.it
%Date: 25 Nov 1993 10:03:38 +0000

%%%%%%%%%%%%%%%%%%%%%%%%%%%%%%%

\magnification\magstep 1
\vsize=8.5 true in
\hsize=6 true in
\voffset=0.5 true cm
\hoffset=1 true cm
\baselineskip=20pt

\bigskip
\bigskip
\bigskip
\vskip 5 truecm
\centerline{\bf DISPERSION LAW OF EDGE WAVES IN THE QUANTUM HALL EFFECT}
\vskip 1.5 truecm

\centerline{S. Giovanazzi, L. Pitaevskii$^*$, and S. Stringari}
\bigskip
\centerline{{\it Dipartimento di Fisica, Universit\'a di Trento,
  I-38050 Povo, Italy}}
\vskip 2 truecm

\par\noindent
{\it{\bf Abstract}. We present a microscopic description
of edge excitations in the quantum Hall effect which is analogous
to Feynman's theory of superfluids.
Analytic expressions for the excitation energies
are derived  in finite dots. Our predictions are  in excellent agreement
with the results of a recent numerical diagonalization.
In the large $N$ limit
the dispersion law is proportional
to $qlog{1\over q}$.
For short range interactions the energy instead behaves
as $q^3$.
The same results are also derived using hydrodynamic theory of incompressible
liquids.}

\vskip 1.0truecm

\par\noindent
\bigskip


\bigskip

\vskip 1.5truecm
$^*$ Permanent address: Kapitza Institue for Physical Problems,
ul. Kosygina 2, 117334 Moscow, Russia.

\par\noindent
\vfill\eject

In the last few years a considerable interest has been devoted to the study
of the edge excitations of 2D charged systems (quantum dots)
in strong magnetic fields [1-5]. These
studies are motivated by the fact that edge
excitations, differently from bulk modes, are gapless
and are consequently particularly relevant for
the thermodynamic behavior at low temperature.
The macroscopic picture underlying many of the available theoretical
works is
the hydrodynamic description of an incompressible 2D liquid characterized by
the propagation of edge waves with drift velocity $v=c{E\over B}$ where
$E$ is the electric field generated by the electrons at the edge of the
droplet (see, for example, ref.[4]).
According to this picture the dispersion should be
$$
\omega = {M \over R} c {E\over B}
\eqno(1)
$$
where $M$ is the angular momentum carried by the wave
and $R$ $\propto \sqrt{N}$ is the radius of the dot ($N$ is the number of
electrons).
The above picture presents however a serious difficulty
since the electric field generated by 2D charged
clusters exhibits a logaritmic
enhancement at the border.
Moreover  considerable theoretical effort has been recently devoted
 to models where electrons interact through effective short range
forces [5] and where consequently the concept of electric field cannot
be used.

The purpose of this work is to give an answer to this problem and
to provide an explicit formula for the dispersion law
employing either a microscopic approach based on Feynman's theory and a
macroscopic description  based on classical hydrodynamics. We will also discuss
the crucial role played
by the neutralizing background in ensuring the stability of
the system.

In the following we will consider the Hamiltonian
$$
H = \sum_k{1\over 2m} ({\bf p}_k + {e\over c}{\bf A}({\bf r}_k))^2
+ e^2 \sum_{k<p} {1\over \mid {\bf r}_k-{\bf r}_p\mid}
-e^2 \sum_k\int \rho_{ion}({\bf r}){1\over
\mid {\bf r}_k-{\bf r}\mid}d{\bf r}
\eqno(2)
$$
where the vector potential has the form
$A_x={1\over 2}yB, A_y=-{1\over 2}xB$, the second term is the e-e
Coulomb interaction, while the last term accounts for the Coulomb interaction
with the neutralizing background for which we make the simple choice
$\rho_{ion}(r) = \rho_0\Theta(r-R)$.
Due to charge neutrality the
electron density in the interior of the dot coincides with the
ion density $\rho_0$. The radius $R$ is
then fixed by the normalization condition
$\rho_0\pi R^2 = N$.

We will focus on the case of integer filling ($\nu=1$)
where it is possible to derive important results
in an analytic way. This fixes a relationship between the
density $\rho_0$ and the
magnetic field.
We also assume that
the magnetic field is strong enough that mixing of states in higher
Landau levels by the Coulomb interaction can be neglected.
An important consequence of the integer filling is that the
ground state of the Hamiltonian (2) is
a Slater determinant [6] built up with the single particle states
$$
\psi_l({\bf r}) =
{1\over \sqrt{2\pi\ell^2 }} {1\over \sqrt l!} ({z\over \sqrt2 \ell})^l
e^{-r^2/4 \ell^2}
\eqno(3)
$$
where $l=0,1, ...,N-1$ and $\ell=(\hbar c/eB)^{1/2}$ is the magnetic length.
This Slater determinant carries
angular momentum $L_0=N(N-1)/2$ and is the state with the lowest energy
because of the presence of the confining
potential. In the absence of such a field the electrons would in fact
occupy states
with higher values of $l$ in order to minimize the Coulomb
repulsion.

The electron density  corresponding to the ground state
is characterized, for large $N$,
by a constant value
$\rho_0 = {1\over 2\pi \ell^2}$ in the interior of a circle of radius
$R=\ell\sqrt{2N}$ and by an edge thickness of the order of the magnetic length
$\ell$.
The electric field
generated by the electrons can be easily calculated for large $N$.
At the border of the cluster we find the result
$E(R) = {e\over 2\pi\ell^2} \log(\beta N)$
with $\beta = 16.4$.
This equation explicitly shows the anticipated logaritmic enhancement.

Our approach to the study of the dispersion of the edge excitations
is based on the idea, currently
considered in the literature [5-7], that the lowest
edge state $\mid M>$ carrying angular momentum $L=L_0+M$
is naturally excited by the collective operator
$$
S^{\dagger}_M
= 2^{-M/2}\sum_{k=1}^N({z_k\over 2\ell }
-2\ell{\partial\over \partial z^*_k})^M =
\sum_i\sqrt{{(i+M)!\over i!}}c^{\dagger}_{i+M}c_i
\eqno(4)
$$
where $c_l$ and $c_l^{\dagger}$ are the electron annihilation and creation
operators relative to the  single particle states (3).
The operator (4) corresponds to the projection of the usual multipole operator
onto the lowest Landau level. It is the edge analog of the
projected density operator used in ref.[8] to study bulk excitations in
the fractional quantum Hall effect.

The key point of the work consists of the ansatz
$\mid M> = S^\dagger_M\mid 0>$.
This ansatz has the form of the approximation employed by Feynman
to describe the density excitations
of superfluid $^4He$. The same method has been succesfully applied
to study bulk excitations in the fractional quantum Hall effect [8].

For large $N$ the transition density $\rho_{tr}({\bf r}) =
<M\mid \sum_k\delta({\bf r}-{\bf r}_k)\mid 0>$
associated with the state $\mid M>$ takes the typical form
$\rho_{tr}({\bf r}) \propto \rho^{\prime}(r) exp(-i\theta M)$
of an edge wave with wave vector $q=M/R$.

The excitation energy of the "Feynman" state can be written in the
following form:
$$
\epsilon(M) = {<M\mid H \mid M> \over <M\mid M>} - <0\mid H\mid 0> =
{<0\mid [S_M,[H,S^{\dagger}_M]]\mid 0> \over <0\mid[S_M,S^\dagger_M]\mid 0>}
\eqno(5)
$$
where we have used the fact that the operator $S_M$ annihilates
the ground state. For a fixed value of $M$ this bound is expected
to coincide with the exact dispersion law
in the large $N$ limit.

In the following we calculate the numerator of eq.(5)
using the Hamiltonian (2). The kinetic term does not
contribute to the the commutator since $S^{\dagger}_M$ does
not excite higher Landau levels.
In order to calculate the double commutator we find it convenient to use
the Tamm-Dancoff relation (see, for example, ref.[9])
$$
<0\mid [S_M,[H,S^{\dagger}_M]]\mid 0> = \sum_{minj}S^*_{mi}A_{minj}S_{nj}
\eqno(6)
$$
where $A_{minj}=\delta_{ij}\delta_{mn}(\epsilon_m-\epsilon_i) +
V^{e-e}_{mjin}$ and $\epsilon_m = \sum_{j=0}^{N-1}
V^{e-e}_{mjmj} + V^{ext}_{mm}$ are the
single particle Hartree-Fock energies written in terms of the usual
two-body and one-body matrix elements
$V^{e-e}_{mjin}$ and $V^{ext}_{mm}$. The quantity
$S_{mi} \simeq \delta_{M,m-i}(N^M)^{1/2}$ is the matrix element of the
operator $S_M$ between
the ground state and the particle-hole state $c_m^{\dagger}c_i\mid 0>$.
It is worth noticing that result (6) is an exact one since the ground state
is a Slater determinant.

Let us first calculate the contribution to the double
commutator (6) arising from
the e-e interaction.
By using non trivial relationships involving matrix elements
of the e-e interaction, we find for $N>>M$ the  most important result
$$
<0\mid [S_M,[V^{e-e},S^{\dagger}_M]]\mid 0> =  M^2[F_N(M)-F_N(1)]
\eqno(7)
$$
whith the quantity $F_N(M)$  defined by
$$
F_N(M) =
e^2\int d{\bf r}_1d{\bf r}_2\psi^*_{N+M}({\bf r}_1)\psi^*_{N-M}({\bf r}_2)
{1\over \mid {\bf r}_1-{\bf r}_2\mid}\psi_N({\bf r}_1)\psi_N({\bf r}_2) \, .
\eqno(8)
$$

Due to the $1/r$ behavior of the Coulomb force
the function $F_N(M)$ exhibits a
${1\over  \sqrt{N}} logN$ dependence. However
the $logN$ term vanishes
in the difference (7) which then
behaves as $N^{-1/2}$:
$$
F_N(M)-F_N(1)
= -{2e^2 \over \ell \pi \sqrt{2N}} \sum_{l=2}^M {1\over 2l-1}
\eqno(9)
$$

The contribution of the confining potential
can be also calculated in the large $N$ limit. With our choice for
$\rho_{ion}$ it is possible to relate this contribution
to the electric field generated by the electrons and we find
$$
<0\mid [S_M,[V^{ext},S^{\dagger}_M]]\mid 0> =
N^M M^2  {e^2 \over \ell 2\pi \sqrt{2N}} log (\beta N)
\eqno(10)
$$

Equations (7-10), together with the result
$<0\mid [S_M,S^{\dagger}_M]\mid 0> = MN^M$,
allow us to
calculate the
excitation energy of the edge modes through the Feynman's relation (5).
The energy can be written as the sum of the two contributions:
$$
\epsilon^{e-e} = - {2Me^2\over \pi\ell\sqrt{2N}} \sum_{l=2}^M {1\over 2l-1};
\ \ \ \ \ \ \ \epsilon^{ext} = {Me^2 \over 2\pi \ell\sqrt{2N}} log(\beta N) \,
{}.
\eqno(11)
$$
The quantity $\epsilon^{e-e}$  is always negative
and vanishes in the $M=1$ mode as a consequence
of the translational invariance of the e-e interaction. For $M=2$
it coincides with the result recently found in ref.[6]. Notice that the
sum of the two terms (11) must be positive in order to ensure the stability
of the ground state.

An important property revealed by eq.(11) is that the e-e term does not
depends linearly on $M$
in contradiction with the eq.(1). Actually for large $M$ we find
$\epsilon^{e-e}_N(M) = -{Me^2\over \pi \ell\sqrt{2N}} log(\gamma M)$
with $\gamma = 0.96$.
This logaritmic behavior is the consequence of the Coulomb interaction as
revealed by eqs.(7-9). Use of the short range
force  $a\nabla^2\delta({\bf r}_i-{\bf r}_j)$ yields
a different dependence:
$$
\epsilon^{sr}= {a \over 4\pi \sqrt\pi \ell^4} N^{-3/2}M(1-M^2)
\eqno(12)
$$

Our predictions for $\epsilon^{e-e}$ and $\epsilon^{sr}$
turn out to be in excellent agreement with the results
obtained in ref.[5] through a numerical diagonalization
of the e-e interaction (see figure 1). In fig.2 one clearly sees the
convergency of the results of ref.[5] to the asymptotic expressions (11)
and (12) holding for $N>>M$.

For large $M$ it is natural to introduce the parametrization
$M=qR=q\ell\sqrt{2N}$
and the dispersion law, given by the sum of the two contributions
(11) takes the form
$$
\epsilon(q)= \epsilon^{e-e} + \epsilon^{ext} ={e^2\over \pi}qlog{q_0\over q}
\eqno(13)
$$
with $q_0=\sqrt{\beta}/\sqrt2\ell\gamma = 3.0/\ell$. Note that in the sum (13)
the $log N$ terms arising from $\epsilon^{e-e}$  and $\epsilon^{sr}$
cancel each other.

For the short range force the dispersion law instead becomes
$$
\epsilon^{sr}(q) = -{a\over \ell \pi\sqrt{2\pi}}q^3 \, .
\eqno(14)
$$
The applicability of both results (13) and (14) is fixed by the long wavelength
condition $q\ell<<1$.

In the last part of the work we show that results (13-14) can be directly
obtained
using classical hydrodynamic theory.
We treat our system as a charged incompressible liquid
($div {\bf v}=0$) characterized by a velocity potential of the form
$\chi = \chi_0e^{iqy-qx-i\omega t}$
where $x$ and $y$ are the directions orthogonal and parallel to the border
respectively (the liquid occupies the plane $x\ge 0$).
This implies $v_y=-iv_x$ and the Euler equation
${\partial {\bf v}  \over \partial t} = -{e\over mc}{\bf v x B} -
{1\over m \rho_0} {\bf \nabla}p^{\prime}$
takes the form
$$
p^{\prime} = -i m\rho_0(\omega+\omega_c){v_x\over q} \, .
\eqno(15)
$$
In eq.(15) $\omega_c=eB/mc$ is the cyclotron frquency and $p^{\prime}$
is the oscillating part of the 2D pressure.
The excess of pressure $p^{\prime}$ is compensated at the border
by a restoring force produced by the electric field. This force
can be calculated by varying the energy of the electric field
$U=1/8\pi \int E^2dV$ with respect to the displacement $\xi(y)$ of the
border of the liquid. Thus the boundary condition takes the form
$p^{\prime}_{_{x=0}}=-\delta U/\delta \xi$.
The calculation of $U$ is drastically simplified if one notices that the main
contribution to the integral originates from the region
$q_0^{-1}<r<q^{-1}$
where
$r$ is the distance from the border and $q_0^{-1}$ is a cutoff length of the
order of the
widht of the border. In this region the deformed edge
can be considerd as a wire of charge density $e\xi(y) \rho_0$ generating
the electric
field $E=2e\xi(y)\rho_0/r$.
Integration of the electric energy in the $xz$ plane then gives
$$
U=e^2\rho_0^2log{q_0\over q}\int \xi^2(y)dy
\eqno(16)
$$
showing in a clear way the physical origin of the logaritmic term.
Since the $x$-component of the velocity of the fluid at the border is given
by $v_x=\partial \xi/\partial t$, we
finally obtain the boundary condition
$p^{\prime}_{_{x=0}} = -2ie^2\rho_0^2{v_x\over \omega} log{q_0\over q}$.
The Euler equation (15) then gives rise to the dispersion law
$$
\omega(\omega+\omega_c)=2{e^2\over m}\rho_0q log{q_0\over q} \, .
\eqno(17)
$$
Equation (17) yields, for low $q$ and $\omega$, the result
$\epsilon(q) = \nu {e^2\over \pi}q log{q_0\over q}$
where we have used the relation $\rho_0 = \nu {eB \over 2\pi c\hbar}$
defining the filling factor $\nu$. When $\nu = 1$
the hydrodynamic formula coincides with
the microscopic result (13).
This $qlog{1\over q}$ dependence has been alreday derived in ref. [10-11].

It is finally interesting to recover result (14) for short range forces.
In this case  the relevant restoring force originates from the surface energy
$U= {\alpha\over 2} \int({\partial \xi(y) \over \partial y})^2 dy$
where $\alpha$ is the surface tension. The resulting
dispersion becomes:
$$
\omega(\omega+\omega_c) = {\alpha  \over m \rho_0} q^3
\eqno(18)
$$
yielding, after identifying $\alpha =
-a\rho_0 /\pi\sqrt{2\pi}\ell^3$, result (14) in the low $q$ regime.
Note that in the figures we have reported the results of ref.[5] with
$\alpha < 0$ corresponding to negative excitation energies.

The exact equivalence between the microscopic and macroscopic results
discussed in this work confirms in a clear way
the general statement that
electrons in a strong magnetic field exhibit a behavior typical of Bose
superfluids and that consequently their dynamics
is properly decribed by the equations of classical hydrodynamics [12].
Both in the Coulomb and short range cases the dispersion law however differs
from the linear law $\epsilon=qv$ currently considered in the literature
and reveals new interesting features exhibited by edge excitations
in the quantum Hall effect.

\bigskip

We wish to thank R. Ferrari for many useful discussions and
R.L. Schult for providing us with his numerical results. L.P. likes
to thank the hospitality of the Department of Physics and the financial support
from the Centre ECT$^*$ at the University of Trento.

\vfill\eject

\par\noindent FIGURE CAPTIONS

\bigskip

Fig.1. Lowest excitation energies $\epsilon^{e-e}$ and
$\epsilon^{sr}$  for $N=400$.
Crosses are taken from ref.[5],  while dots are the predictions
of eqs.(11) and (12). The units are $e^2/l$ and
$10^{-3}a/l^4$ for the Coulomb and short-range interactions respectively.

\bigskip

Fig.2. $N$-dependence of the lowest excitation energies
$\epsilon^{e-e}$ and
$\epsilon^{sr}$ for $M=10$
Crosses are taken from ref.[5],  while the dashed lines are the asymptotic
predictions (11) and (12). For the units see fig.1.

\bigskip

\par\noindent REFERENCES

\bigskip

\item{1.} B.I. Halperin, Phys.Rev. B {\bf 25}, 2185 (1982);

\item{2.} M. Stone, Ann.Phys. (NY) {\bf 207}, 38 (1991);

\item{3.} A.H. MacDonald, Phys.Rev.Lett. {\bf 64}, 220 (1990);

\item{4.} X.G. Wen, Phys.Rev. B {\bf 43}, 11025 (1991);

\item{5.} M. Stone, H.W. Wyld and R.L. Schult,
Phys.Rev. B {\bf 45}, 14156 (1992);

\item{6.} A.H. MacDonald, S.R.E. Yang and M.D. Johnson, Aust. J. Phys.
{\bf 46}, 345 (1993);

\item{7.} M. Marsili, Phys.Rev. B, in press;

\item{8.} S.M. Girvin, A.H. MacDonald and P.M. Platzman, Phys.Rev.Lett.
{\bf 54}, 581 (1985); Phys.Rev. B {\bf 33}, 2481 (1986);

\item{9.} D.J. Rowe, {\it Nuclear Collective Motion}, (Methuen, London,
1970);

\item{10.} V.A. Volkov and S.A. Mikhailov, JETP Lett. {\bf 42}, 556 (1985);

\item{11.} X.G. Wen, Phys. Rev. B {\bf 44}, 5708 (1991);

\item{12.} M. Stone, Phys.Rev. B {\bf 42}, 212 (1990).

\bye